\documentstyle[12pt,epsf]{article}

\def\Journal#1#2#3#4{{#1} {\bf #2}, #3 (#4)}

\def\NPB{{\em Nucl.\ Phys.}\ B}
\def\PLB{{\em Phys.\ Lett.}\  B}
\def\PRL{\em Phys.\ Rev.\ Lett.\ }
\def\PRD{{\em Phys.\ Rev.}\ D}

\def\beq{\begin{equation}}
\def\eeq{\end{equation}}
\def\beqa{\begin{eqnarray}}
\def\eeqa{\end{eqnarray}}

\begin{document}

\begin{flushright}
ITP-SB-99-27
\end{flushright}

\begin{center}
{\bf RESUMMATION FOR HEAVY QUARK PRODUCTION\\ NEAR PARTONIC THRESHOLD}
\footnote{Talk presented at 
Workshop on Physics With Electron Polarized Ion Collider, EPIC '99,
April 8-11, 1999, Indiana University Cyclotron Facility.}

\vbox{\vskip 0.25 true in}

Gianluca Oderda${}^a$, Nikolaos Kidonakis${}^b$, George Sterman${}^a$

\vbox{\vskip 0.25 true in}

{\it ${}^a$Institute for Theoretical Physics\\ State University of New
York at Stony Brook\\
Stony Brook, NY 11794-3840} 

\medskip

{\it ${}^b$Department of Physics\\ Florida State University,
Tallahassee, FL 32306-4350}
\end{center}

\begin{abstract}
We review some techniques of
resummation applied to heavy quark production
in hadronic scattering and electroproduction near partonic
threshold, and discuss the reduction
of factorization
scale dependence in resummed cross sections.
\end{abstract}
\section{Introduction}

In this talk, we will discuss high-order corrections
to heavy quark production in QCD.
To get a better confidence in our ability to
predict cross sections for various heavy quark
reactions,  including open flavor and quarkonia production, we
need to gain  control
over a class of corrections
associated with what is sometimes called partonic threshold.
This talk will describe 
the nature of these corrections, and report on 
the progress of the past few years \cite{ksnp,bcmn,kos,kidvog,1pI,nkrv}
in their resummation to all 
orders in perturbation theory.

Corrections due to resummation may turn out to be small, in which case
our confidence in low-order perturbative cross sections should
increase, or, they may turn out to be large, 
and may afford sensitivity to QCD dynamics in 
a regime where all orders of perturbation theory
are relevant.  It is possible that both scenarios apply in 
different cross sections, and/or in different kinematic ranges
of the same cross section.

\section{Threshold Resummation for Hard Inclusive Scattering}

Inclusive heavy quark production is just one example 
of a QCD process at 
large momentum transfer.  
We may consider the class of cross sections 
in which we sum over all final states 
that include any heavy system $F=Q\bar Q, \dots$, 
which can only be  produced by a short-distance
process in partonic scattering.  
   
\begin{figure}[ht]
\mbox{\hskip 2.5 cm \epsfysize=6cm \epsfxsize= 6.8 cm \epsfbox{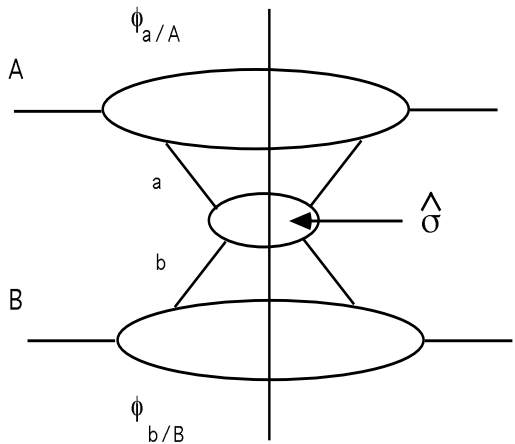}}
\caption{Hard-scattering cross section in cut (unitarity) diagram
notation.}
\label{cutcross}
\end{figure}

Suppose for simplicity that the total mass $Q$ of the system $F$
is of order $\sqrt{S}$, the total (hadronic) center of mass energy,
and that the rapidity $y$ of the produced system is not a large
parameter.  Such a cross section can be expressed
to leading power by the factorized expression
\footnote{The extension to single-particle inclusive
cross sections has recently been carried out. {\protect \cite{1pI}}}
\beqa
{d\sigma_{AB\rightarrow FX}\over dQ^2dy}
&=&
\sum_{ab}
\int_{Q^2/S}^1dz\;
\int{dx_a} {dx_b}\; \phi_{a/A}(x_a,\mu^2) 
\phi_{b/B}(x_b,\mu^2) \nonumber \\
&\ & \times \delta\left(z-{Q^2 \over x_ax_bS}\right)\; 
\, 
{\hat \sigma}_{ab\rightarrow FX}\left (z,{\rm e}^{-2y}{x_a\over x_b},
{Q\over\mu},Q,
\alpha_s(\mu^2)\right)\, ,
\label{basicfact}
\eeqa
which is illustrated in Fig.\ \ref{cutcross}.   
The $\phi$'s are
parton distributions (in some factorization
scheme, like DIS or $\overline{\rm MS}$), while
$\hat{\sigma}$ is a partonic hard-scattering function, which
at lowest order is the Born cross section for
$a+b\rightarrow F+X$,
\beq
{\hat\sigma}=\sigma_{\rm Born}+{\alpha_s\over\pi}{\hat\sigma}^{(1)}
+\dots\, .
\label{sigmaborn}
\eeq
These relations may be
written down for both unpolarized and polarized cross sections.
Corrections in either case are power-suppressed \cite{qs}; beginning at
${\cal O}(1/Q^2)$ for unpolarized or longitudinally polarized
cross sections, and at ${\cal O}(1/Q)$ for transversely polarized
cross sections.

In Eq.\ (\ref{basicfact}), 
the limit $Q^2/x_ax_bS\rightarrow 1$ is associated with corrections
that tend to enhance the cross section.  These enhancements 
are the object of threshold resummation.  We will 
discuss below the origin of this terminology, why such corrections
can sometimes be large, and how they can be resummed to all orders.
Much of the following discussion follows Ref.\ \cite{laThuile}.

The kinematics of the partonic process require that 
$x_ax_bS\ge Q^2$, so that $z\le 1$
in Eq.\ (\ref{basicfact}).  At $z=1$,
the partons have just enough energy to produce the observed final
state, with no extra hadronic radiation.  This is what we
shall refer to as the elastic limit \cite{cls}, or partonic threshold.
It is important to distinguish partonic threshold from the usual concept of
a threshold.  In particular, in heavy quark production, we shall
assume that the heavy quarks of mass $M_Q$ are produced with nonzero velocity
$\beta$, and hence with a total invariant mass $Q^2>4M_Q^2$.  Thus,
only for $\beta=0$ does partonic threshold coincide with 
true threshold.  For the Drell-Yan production of 
highly relativistic lepton pairs
with $Q^2\gg 4m_\ell^2$,
partonic threshold still refers to $z=1$, and
is a source of potentially large corrections.

Typical hard-scattering cross sections $\hat\sigma(z\dots)$ 
in Eq.\ (\ref{basicfact}) are distributions 
in the variable $z$ rather
than simply functions of $z$, because they include contributions from
virtual as well as real gluons. 
We are interested in a class of potentially large, positive corrections 
due to such distributions, which
occur in all $\sigma^{(n)}$.  We next explain in what sense they are
``large", and why they are positive to all orders.

At order $\alpha_s^n$, the leading logarithmic distributions in Eq.\
(\ref{basicfact}) are of
the form (suppressing color factors, which depend on parton type)
\beq
-{\alpha_s^n\over n!}\; \left [ {\ln^{2n-1}\left((1-z)^{-1}\right)\over 1-z} \right ]_+\, ,
\label{lln}
\eeq
whose integral with a smooth function ${\cal F}(z)$ (such as the convolution
of parton distributions in Eq.\ (\ref{basicfact})) is defined as
\beqa
-{\alpha_s^n\over n!}\int_0^1dz\; 
{{\cal F}(z)-{\cal F}(1)\over 1-z}\; \ln^{2n-1}\left((1-z)^{-1}\right) &\ &
\nonumber\\
&\ & \hspace{-6.8 cm}
= {\alpha_s^n\over n!}\int_0^1dz\; {\cal F}'(1) \ln^{2n-1}\left((1-z)^{-1}\right)+\dots
\sim {\alpha_s^n\over n!}{\cal F}'(1){(2n-1)!}+\dots\, ,
\label{factorial}
\eeqa
where we have kept only the first term in the expansion of ${\cal F}(z)$ 
about $z=1$.  It is evident that such terms give, at least
formally, contributions that grow even faster than $n!$ at $n$th order.
If they had alternating signs,
these contributions might add up to a finite number somehow, but
they are all of the same sign.  

\begin{figure}[ht]
\mbox{\hskip 1.25 cm \epsfysize=4.5cm \epsfxsize=9cm \epsfbox{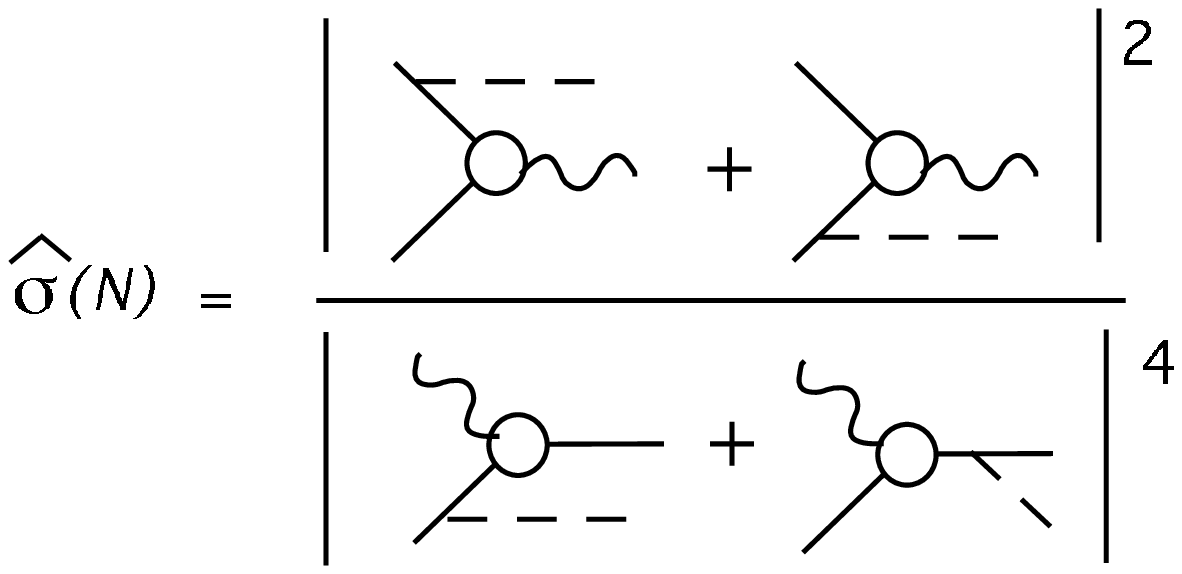}}
\caption{Schematic representation of moments of the
Drell-Yan partonic hard-scattering function.}
\label{mtratio}
\end{figure}

The signs in Eq.\ (\ref{factorial}) come directly from the manner in which
hard-scatter-ing cross sections are computed.  
The fully inclusive Drell-Yan cross section $d\sigma/d Q^2$ illustrates the
situation.
The convolution in
Eq.\ (\ref{basicfact}) factors into simple products 
of functions for each pair of initial-state partons $a$ and $b$, under
moments with respect to $\tau=Q^2/S$,
\beqa
\sigma^{\rm DY}_{AB}(N,Q)
&=&
\int_0^1 d\tau \tau^{N-1}\; {d\sigma_{AB}^{\rm  DY}(\tau,Q)\over dQ^2}
\nonumber\\
&=& 
\sum_{ab} \phi_{a/A}(N,\mu)\;
\hat{\sigma}_{ab}^{\rm DY}(N,Q,\mu)\;
\phi_{b/B}(N,\mu)\, .
\eeqa
Identifying $\sigma_{AB}^{ab}$ as the contribution
to the cross section from the parton combination $a+b$, we find 
\beq
\hat{\sigma}_{ab}^{\rm DY}(N,Q,\mu) 
= 
\prod_{i=a,b}
{1\over\left[\int_0^1dx\; x^{N}\; \phi_i(x,\mu)\right]}\, 
\int_0^1 d\tau\; \tau^{N-1}{d\sigma^{ab}_{AB}(\tau)\over dQ^2}\, .
\eeq
  Neglecting parton labels, this is the ratio of the moment of a 
physical cross section
to a product of moments of parton distributions.  Because $\hat{\sigma}$
is, by construction, dependent only on short-distance behavior, 
the ratio may be computed in perturbation theory, as illustrated
schematically in Fig.\ \ref{mtratio} (for the DIS scheme).   
The numerator is a moment of the perturbative
partonic Drell-Yan cross section,
while the denominator is the product of 
moments of two perturbative parton distributions.  For quark-antiquark
processes, the parton distributions are the same, so the denominator
is the square of squared partonic amplitudes, summed over final states.

At each order, both the numerator and denominator in Fig.\ \ref{mtratio} have
double-logarithmic terms like Eq.\ (\ref{lln}).
Before moments, the perturbative Drell-Yan cross sections include 
distributions in $1-z$, like Eq.\ (\ref{lln}) above,
 while in the deeply inelastic scattering cross section
the same sort of terms depend 
on Bjorken $x$ through $1-x$.  After moments, both give double-logarithmic 
$\alpha_s^n\ln^{2n}N$
at $n$th order, with $N$ the moment variable.
These leading
logarithms are the finite remainders of
corrections from $n$ pairs of
 real and virtual gluons
that attach to the scattered quarks in DIS and the annihilating
pair in DY. 
In the denominator, each DIS parton distribution,
which is itself of the form of a cross section, has
both incoming {\it and} outgoing
quarks, while in the numerator, DY involves incoming quarks only.

On physical grounds, the
numerator must vanish as $z\rightarrow 1$ 
and the denominator must vanish as $x\rightarrow 1$.
 The reason for this
may be seen by recalling the relations of $z$ and $x$ to the invariant
mass $W$ of hadrons in the final state for the two cases:
\beqa
{\rm DY}&:& W^2\le {1\over 4}Q^2(1-z)^2\, ,\nonumber \\
{\rm DIS}&:& W^2\sim Q^2(1-x)/x\, .
\label{wsquared}
\eeqa
The limits $z\rightarrow 1$ and $x\rightarrow 1$ thus both
correspond to nearly elastic scattering \cite{cls}: for Drell-Yan, the
annihilation of a quark pair into an electroweak vector boson,
for DIS, the scattering of a quark into a nearly massless jet
of particles.   
It is not difficult to verify that in these limits, the partonic
cross section is suppressed \cite{oldDY}.
Indeed, in these
limits, we expect the coherent scattering of hadronic bound states,
whose contributions are normally suppressed by a power of $Q$
compared to incoherent partonic scattering, to dominate \cite{bl}.
The outgoing quarks give an extra
suppression in DIS when the  
hard-scattering function $\hat{\sigma}$, Fig.\ \ref{mtratio},
is computed in perturbation theory.
That is, the product of DIS denominators is suppressed
even more than the single DY numerator.  Then the ratio grows
with moment $N$, from the elastic limit in $z$ space.
This is the source of the terms shown in Eq.\ (\ref{lln}) in
$\hat{\sigma}$, and is the reason why they all have the same sign.

The Drell-Yan cross section is the benchmark for
the resummation of singular distributions.  As above, singular
distributions at $z=1$ translate into logarithms of the
moment variable $N$. 
 Logarithms of $N$ (to all logarithmic
order, not just leading or next-to-leading logarithm) 
in the moments of the inclusive Drell-Yan cross section 
exponentiate \cite{cls,oldDY},
\beq
\hat{\sigma}_{q\bar q}^{\rm DY}(N,Q,\mu)=\sigma_{\rm Born}(Q)\;
e^{C(\alpha_s)+E(N,Q,\mu,\alpha_s)}\, ,
\label{dyexp}
\eeq
where $\alpha_s$ stands for $\alpha_s(\mu^2)$.  In the exponent,
the function $C$ is known to two loops, while the function $E$,
which organizes all logs of $N$, has the following form in 
the $\overline{\rm MS}$ scheme,
\beqa
E(N,Q,\mu,\alpha_s)
&=&
-\int_0^1dx\; {x^{N-1}-1\over 1-x} \bigg [
\int_{(1-x)^2Q^2}^{\mu^2}{d\mu'{}^2\over \mu'{}^2}\;
g_1\left(\alpha_s[\mu'{}^2]\right)\nonumber\\
&\ & \quad\quad + g_2\left(\alpha_s\left[Q^2\right]\right)
\, \bigg]\, .
\label{dyexponent}
\eeqa
The functions $g_1$ and $g_2$ are finite series in $\alpha_s$ \cite{cls},
given for $\overline{\rm MS}$ by
\beq
g_1(\alpha_s)= 2C_F
\left ( {\alpha_s\over \pi} + {1\over 2}K\; 
\left({\alpha_s\over \pi}\right)^2\right )+\dots\, ,\quad
g_2(\alpha_s) = {\cal O}(\alpha_s^2)\, ,
\label{g12def}
\eeq
where \cite{kt}
\beq
K=C_A\left({67\over 18}-{\pi^2\over 6}\right ) -{5\over 9}n_f\, .
\eeq
Eqs.\ (\ref{dyexp}) and (\ref{dyexponent}) resum all
logarithms of $N$ in the sense of an order-by-order
expansion, by reexpanding the running couplings in
terms of $\alpha_s$.  The resummed integrals, however,
are ill-defined for $x\rightarrow 1$, no matter how large
$Q^2$ is, since the perturbative 
running coupling diverges at $\mu'{}^2=\Lambda^2$.  
Such a divergence is sometimes called an infrared renormalon,
and may give hints on the structure of power corrections
to Eq.\ (\ref{basicfact}).  Most importantly, it is possible
to regulate the singularity in the running coupling
without affecting the leading power behavior of the resummed
cross section \cite{ams}.  Different approaches to this problem
can give somewhat different predictions \cite{lsvn,cmnt,bc},
but in any reasonable prescription the threshold
corrections remain relatively modest, although
nonnegligible in at least some physically relevant
regions.

It has been noted
in several phenomenological applications that threshold resummation,
and even fixed-order expansions based upon it,
significantly reduces sensitivity to the factorization scale
\cite{bcmn,kidvog,bc,phoresum,lm,ko}.  The reason
for this reduction is easy to see 
in the dependence on the factorization scale, $\mu$
in Eq.\ (\ref{dyexponent}).  Indeed, for the Drell-Yan,
and all related cross sections, the dependence on
 $\mu$ in the hard-scattering
function matches that
part of the scale dependence of the $\overline{\rm MS}$ 
distributions due to the singular
part of the splitting functions $P_{ii}(z)$.  That is,
$g_1$ in Eq.\ (\ref{g12def})
is exactly the coefficient of $1/(1-z)_+$
in $P_{qq}(z)$.  Because resummation matches phase
space to all orders at partonic threshold \cite{oldDY}, all associated factorization
scale dependence is also matched.  To be specific, we 
may write the moments of the Drell-Yan cross section in a 
form that resums the singular threshold distributions,
\beqa
\sigma^{\rm DY}_{AB}(N,Q)
&=& 
\sum_q\;
 \phi_{q/A}(N,\mu)\;
\hat{\sigma}_{q\bar q}^{\rm DY}(N,Q,\mu)\;
\phi_{\bar q/B}(N,\mu)
\nonumber\\
&=& 
\sum_q\;
\phi_{q/A}(N,\mu)\; M_q(\mu,Q)
\hat{\sigma}_{q\bar q}^{\rm DY}(N,Q,Q)\;
\phi_{\bar q/B}(N,\mu)\; M_{\bar q}(\mu,Q)
\nonumber\\
&\ & \quad \quad \quad  + {\cal O}({1/ N})\, ,
\eeqa
where in the second line the factors $M_q(\mu,Q)$ absorb the
$\mu$-dependence in the exponentials of Eq.\ (\ref{dyexponent}).
These factors compensate for the singular part of the
evolution of the parton distributions, and the
 $\mu$-dependence of the resummed expression
is suppressed by a power of the moment variable,
\beq
\mu{d\over d\mu}\left[\, \phi_{q/A}(N,\mu)\; M_q(\mu,Q)\; \right]
=   {\cal O}({1/ N})\, .
\eeq   
Of course, the importance of the remaining sensitivity to $\mu$
depends on  the kinematics.     

\section{Resummation with Color Exchange}

\begin{figure}[t]
\mbox{\hskip 1.5 cm \epsfysize=8cm \epsfxsize=9cm \epsfbox{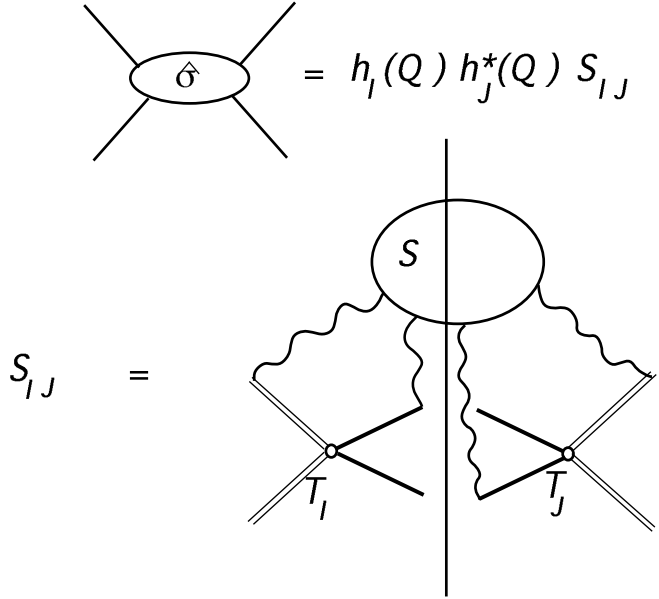}}
\caption{Representation of the factorization
of the hard scattering function $\hat{\sigma}$ 
near the elastic limit.  The second part shows
the soft-gluon matrix $S_{IJ}$ as a
cut diagram for the scattering of incoming ordered
exponentials (double lines - the incoming partons
in the eikonal approximation) to give outgoing
ordered exponentials (bold lines - the outgoing
heavy quarks in the eikonal approximation). For simplicity,
only a few of the possible gluon 
interactions with the ordered exponentials are shown.}
\label{HS}
\end{figure}

Beyond leading logarithms, there are important
differences between the electro-weak-induced Drell Yan
cross sections and the QCD-induced top or jet cross sections.
These are due primarily to the presence of final-state
radiation from scattered quarks in the latter case, which
is absent in the former, and to the interplay
of color exchange in the hard scattering with the soft
radiation.  
As in the case of
Eq.\ (\ref{dyexp}), resummation is
based first of all on the factorization properties
of the cross section in the neighborhood of the
elastic limit \cite{oldDY}.  The situation is illustrated in
Fig.\ \ref{HS}.  
Near the elastic limit, all gluons emitted into the final
state have energies limited by $(1-z)Q\ll Q$.
Correspondingly, gluons with energies of order $Q$
can appear only in virtual diagrams.  Standard
factorization methods may then be used to separate 
the relatively soft but still perturbative
gluons from the underlying hard scattering.  
This may be done 
order-by-order in perturbation theory, keeping 
both the hard and soft factors
of $\hat{\sigma}$ free of soft and 
collinear divergences.  The process of factorization
may be thought of as the construction of an
effective field theory for soft gluons in the
presence of the hard scattering \cite{ksnp,kos}.
 
In the relevant effective field theory, the incoming
partons that annihilate into the heavy quarks and
the outgoing heavy quarks themselves are represented by
ordered exponentials (Wilson lines).  The Wilson lines are 
tied together in the amplitude and its complex conjugate
at local vertices, $T_I$ and $T_J$  
in Fig.\ \ref{HS}, which describe the flow of color 
between the initial and final states.  
Indices $I$ and $J$
label matrices in color space.
We will consider here the annihilation of light
quarks (color indices $a_1$ and $a_2$)
into heavy quarks (indices $a_3$ and $a_4$), 
\beq
q_{a_1}(p_a)+{\bar q}_{a_2}(p_b) 
\rightarrow Q_{a_4}(p_1) + {\bar Q}_{a_3}(p_2)\, ,
\label{qqQQ}
\eeq
with kinematic invariants,
\beq
t_1=(p_a-p_2)^2-m^2, \quad u_1=(p_b-p_2)^2-m^2, \quad s=(p_a+p_b)^2\, .
\eeq
In this case a convenient basis for the $T$'s 
is one that represents
color singlet and octet exchange in the  $s$-channel,
\beq
\left( T_1 \right)_{\{a_i\}} = \delta_{a_1a_2}\delta_{a_3a_4}\, , 
\quad \quad
\left( T_2 \right)_{\{a_i\}} =
\sum_c \left(T_c^{(F)}\right)_{a_2a_1}\left(T^{(F)}_c\right)_{a_4a_3}\, .
\label{Tdefs}
\eeq
Other bases, particularly singlet exchange
in the $s$- and $u$- channels, are also interesting \cite{ksnp,kos}.
As in Fig.\ \ref{HS}, each choice of effective vertices
leads to a separate soft function $S_{IJ}$, which
depends on $(1-z)Q$ only, rather than $Q$ itself.

The (virtual) hard-scattering functions, $h_I(Q)$ and $h^*_J(Q)$, 
 are labelled by
the same color exchange indices.  As in most factorizations
and constructions of effective field theories, the new
 vertices require renormalization.  Thus
we renormalize the soft functions \cite{kos,BottsSt},
\beq
S_{JI}^{({\rm un})}\left((1-z)Q\right)=
Z^\dagger_{JJ'} S_{J'I'}^{({\rm ren})}\left((1-z)Q\right)Z_{I'I}\, ,
\label{Sren}
\eeq
and the hard functions
\beq
h_I^{({\rm un})}(Q)=Z^{-1}_{IJ}h_J^{({\rm ren})}(Q)\, ,
\label{hren}
\eeq
where the $Z_{KL}$ are cutoff-dependent renormalization  constants.
  
Solutions to the renormalization group equation for $S_{IJ}$
that follows from (\ref{Sren}) are generally ordered
exponentials \cite{cls,BottsSt}, but 
at leading logarithm in $S_{IJ}$, which is next-to-leading
logarithm in the overall cross section, we can
diagonalize the anomalous dimension, and separate the
evolution of particular linear combinations of composite color
vertices \cite{nkrv}.  
 
We can apply these results to the production of a pair of heavy quarks
with total invariant mass $Q\ge 2M_Q\equiv 2m$ at fixed rapidity $y$.
The cross section is, as usual, a convolution 
of hard-scattering functions $\hat{\sigma}_{ab}$ with
parton distributions $\phi_{q/A}$ and $\phi_{{\bar q}/B}$,
as in Eq.\ (\ref{basicfact}), with $F=Q{\bar Q}$.
Corresponding to the Drell-Yan result, (\ref{dyexp}), we now have,
to next-to-leading logarithmic accuracy, choosing $\mu=Q$,
\beq
\hat{\sigma}_{q{\bar q}\rightarrow Q{\bar Q}}(N)
=\sum_{IJ}S^{(0)}_{JI}\; h_I(Q)h^*_J(Q)\;
e^{C'_{JI}(\alpha_s)+E_{JI}(N,\alpha_s)}\, ,
\label{qqQQexp}
\eeq
where again $\alpha_s$ stands for $\alpha_s(Q^2)$.  
The function $C'$ is known to one loop only at
this time.  To next-to-leading log,
we need only the lowest-order 
soft functions, $S_{IJ}^{(0)}\sim\delta_{IJ}$.
The function $E_{IJ}$,
which contains the logs of the moment variable
$N$ has a form very similar to the Drell-Yan case, but now
with a dependence on the effective color vertices, through
a third function, $g_3$,
\beqa
E_{JI}^{(ab)}(N,\alpha_s)
&=&
-\int_0^1dx{x^{N-1}-1\over 1-x} \bigg [
\int_{(1-x)^2Q^2}^{Q^2}{d\mu'{}^2\over \mu'{}^2}\;
g_1\left(\alpha_s[\mu'{}^2]\right)
\nonumber\\
&\ & \quad\quad + g_2\left(\alpha_s\left[Q^2\right]\right)
+ g_3^{(I)}\left(\alpha_s\left[(1-x)^2Q^2\right]\right)
\nonumber\\
&\ & \quad\quad 
+ g_3^{(J)}{^*}\left(\alpha_s\left[(1-x)^2Q^2\right]\right)\, \bigg]\, .
\label{qqQQexponent}
\eeqa
As before, the $g_i$, $i=1,2,3$ are finite functions of their
arguments.  In the $\overline{\rm MS}$ scheme, $g_1$ and $g_2$ are
given for incoming light quarks by (\ref{g12def}) above.
Dependence on color exchange in the hard
scattering is contained entirely in 
the new functions $g_3^{(I)}$. 
To determine $g_3^{(I)}$, we
go to a color basis that diagonalizes the
renormalization matrix $Z_{IJ}$ in eqs.\ (\ref{hren})
and (\ref{Sren}).  In this basis,
\begin{equation}
g_3^{(I)}[\alpha_s]=-\lambda_I[\alpha_s]\, ,
\end{equation}
where the eigenvalues $\lambda_I$ 
of the corresponding anomalous dimension matrix
are complex in general, and may depend
on the directions of the incoming and outgoing partons.

The anomalous dimension matrix
of the effective vertices $T_I$ in Fig.\ \ref{HS} for
light to heavy quark annihilation in the 
singlet-octet basis of Eq.\ (\ref{Tdefs}) \cite{ksnp} is:
\beq
\Gamma_{S'}=\frac{\alpha_s}{\pi}\left(
                \begin{array}{cc}
              - C_F(L_{\beta}+1+\pi i)  &  \frac{C_F}{C_A} \ln\left(\frac{u_1}{t_1}\right)  \\
\ & \ \\
                2\ln\left(\frac{u_1}{t_1}\right)    & C_F
\left[4\ln\left(\frac{u_1}{t_1}\right)-L_{\beta}-1-\pi i \right]\\
\ & +\frac{C_A}{2}\left[-3\ln\left(\frac{u_1}{t_1}\right)
-\ln\left(\frac{m^2s}{t_1u_1}\right)+L_{\beta}+\pi i \right]
                \end{array} \right)\, .
    \label{gammaoneloop}
\eeq
Here $L_\beta$ is the vertex function in the eikonal approximation for the 
production of a pair of heavy quarks with center of mass velocity $\beta$,
\beq
L_{\beta}=\frac{1-2m^2/s}{\beta}\left( \ln\frac{1-\beta}{1+\beta}
+i\pi\right), 
\quad \beta=\sqrt{1-4m^2/s}\, .
\eeq
Solving for the eigenvalues, substituting them in Eq.\ (\ref{qqQQexp}),
and expanding the result to first order in $\alpha_s$,
we can derive an explicit one-loop expression for the
cross section for heavy quark production through light
quark annihilation \cite{ksnp,bcmn,kidvog}.  This result is
consistent \cite{1pI} with the explicit one-loop formulas given 
in \cite{mengetal} for the 
single-particle inclusive cross section.    

Much the same considerations apply
to electroproduction of heavy quarks.  It is interesting
to note that
for the process \cite{lm}
\beq
\gamma^*(p_a)+g_i(p_b)\rightarrow 
Q_{a_4}(p_1) + {\bar Q}_{a_3}(p_2)\, ,
\eeq
as for direct photon production \cite{1pI,nkrv,phoresum} there is only
a single color tensor, coupling the produced
pair to the gluon in an octet state.  As a result, there
is only one eigenvalue, which in this case is given
(in the conventions of Ref.\ \cite{kos}) by
\beq
\lambda_8
=
\frac{\alpha_s}{\pi}\left(
{C_A\over 2} \left[ \ln{t_1u_1\over m^2s} - 1 +i\pi \right] + 
({1\over 2}C_A - C_F)\; [L_\beta + 1]
\right)\, .
\eeq
Resolved processes in photoproduction, however,
involve nontrivial color mixing \cite{gap}.
In the limit $\beta\rightarrow 0$, we can
make contact with quarkonia production
in the nonrelativistic QCD (NRQCD) 
formalism \cite{bbl}.

\section{Conclusion}

The use of resummation as a tool for
heavy-quark and other hard-scattering
cross sections may help to put the successes
and shortcomings of fixed-order perturbative
QCD into a larger context.  The reduction of
factorization scale dependence is an encouraging
observation in this direction.  It will
be interesting to study as well
renormalization scale independence \cite{rgscale}.   Other
directions for  investigation include
the inversion of the moments and the closely-related
issue of power-suppressed corrections. 
The development of resummed polarized cross sections,
for which next-to-leading order calculations 
are becoming available \cite{spinhq}, is sure to be 
a further source of insight.

\noindent
\section*{Acknowledgements}

The treatment of scale-independence in this
talk grew out of discussions with Werner Vogelsang. 
We thank Eric Laenen and Jeff Owens for 
their collaboration in
the development of some of this work,  and Stan Brodsky
and Paolo Nason for
helpful conversations.  
This work was supported in part by the National Science Foundation,
under grant PHY9722101, and by the U.S.\ Department of Energy.

\end{document}